\documentclass[11pt,twoside]{article}

\usepackage{asp2006}
\usepackage{epsf}
\usepackage{psfig}
\usepackage{lscape}
\usepackage{graphicx}

\begin{document}
\title{Old stellar counter-rotating components in elliptical-spiral mergers: exploring the GalMer database}   
\author{P. Di Matteo, F. Combes, A. L. Melchior, B. Semelin}   
\affil{LERMA, UMR 8112, CNRS, Observatoire de Paris, 61 Avenue de l'Observatoire, 75014 Paris, France}    

\begin{abstract} 
We investigate, by means of numerical simulations, the kinematics of elliptical-spiral merger remnants. Counterrotation can appear both in coplanar and in non-coplanar retrograde mergers, and it is mostly associated to the presence of a disk component, which preserves part of its initial spin. In turn, the external regions of the two interacting galaxies acquire part of the orbital angular momentum, due to the action of tidal forces.
\end{abstract}


\section{The Simulations}   
The simulations analyzed are a small subset of the simulations realized in the framework of the GalMer Project \citep{dimatteo07a, dimatteo07b}. 
\section{Coplanar mergers: 2D velocity maps and rotation curves}
\begin{figure}
 \begin{minipage}[b]{6cm}
   \centering
  \includegraphics[width=5.8cm,angle=0]{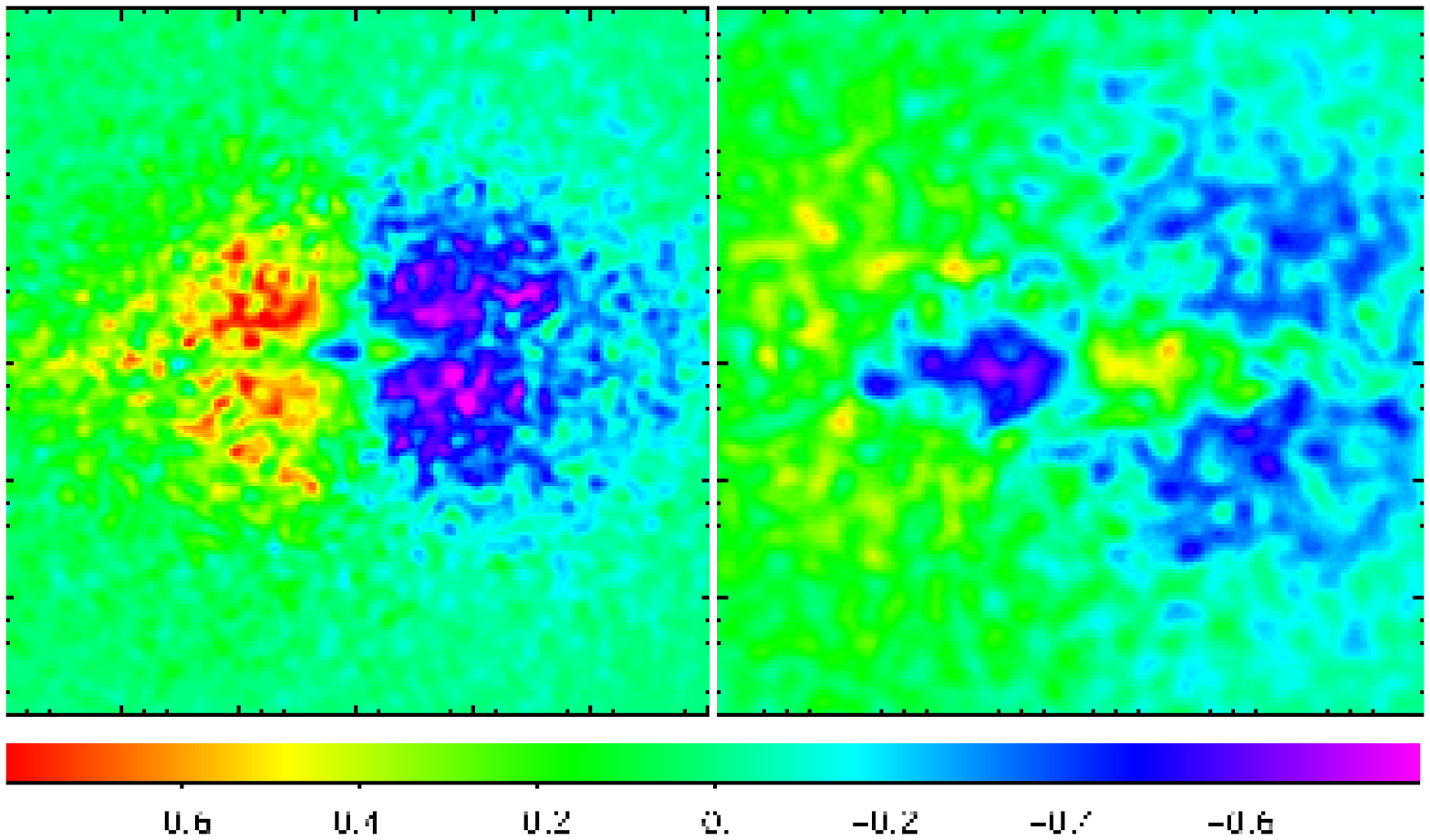}
 \end{minipage}
\begin{minipage}[b]{6cm}
   \centering
  \includegraphics[width=5.8cm,angle=0]{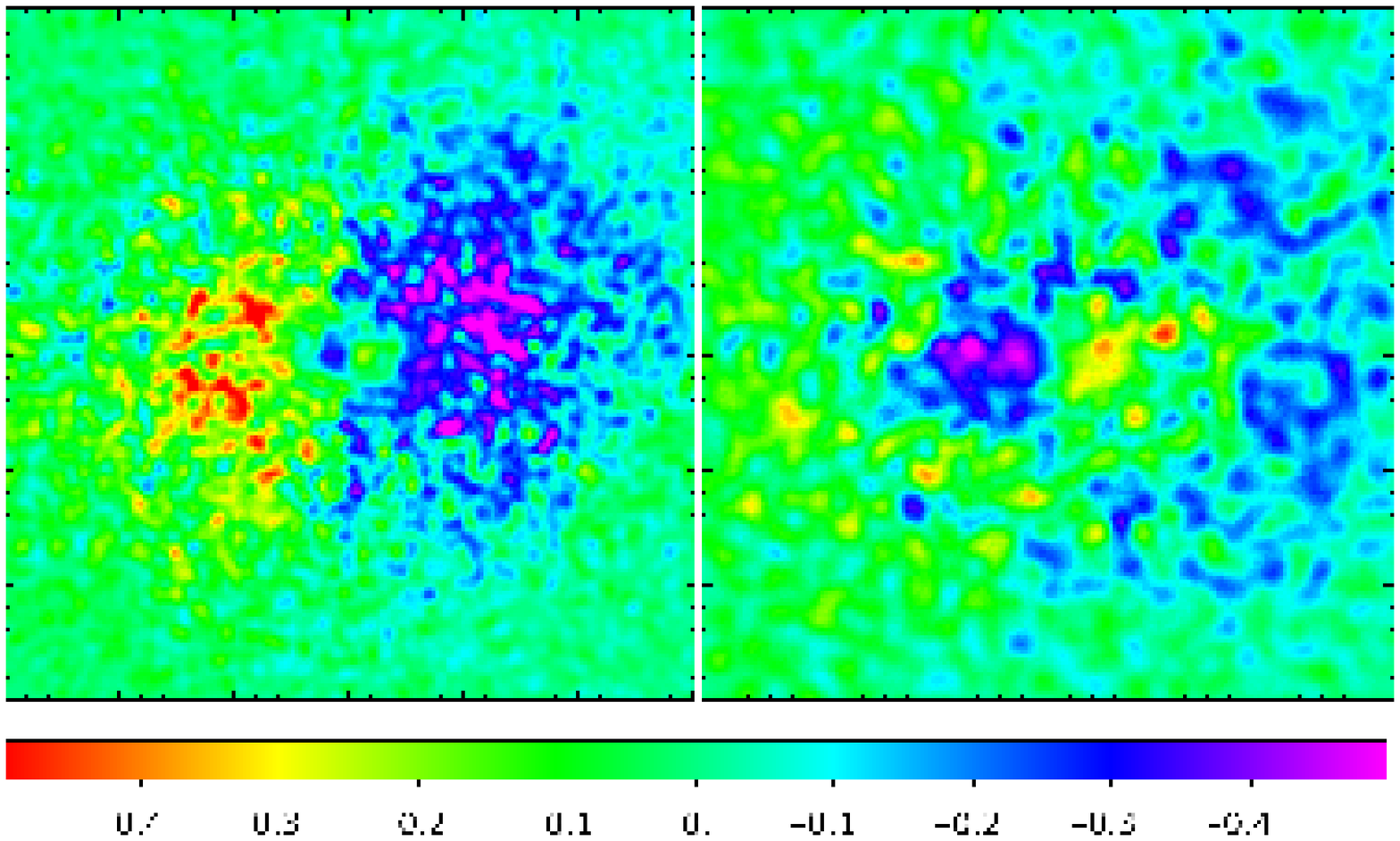}
 \end{minipage}
\\
\begin{minipage}[b]{6cm}
   \centering
  \includegraphics[width=5.8cm,angle=0]{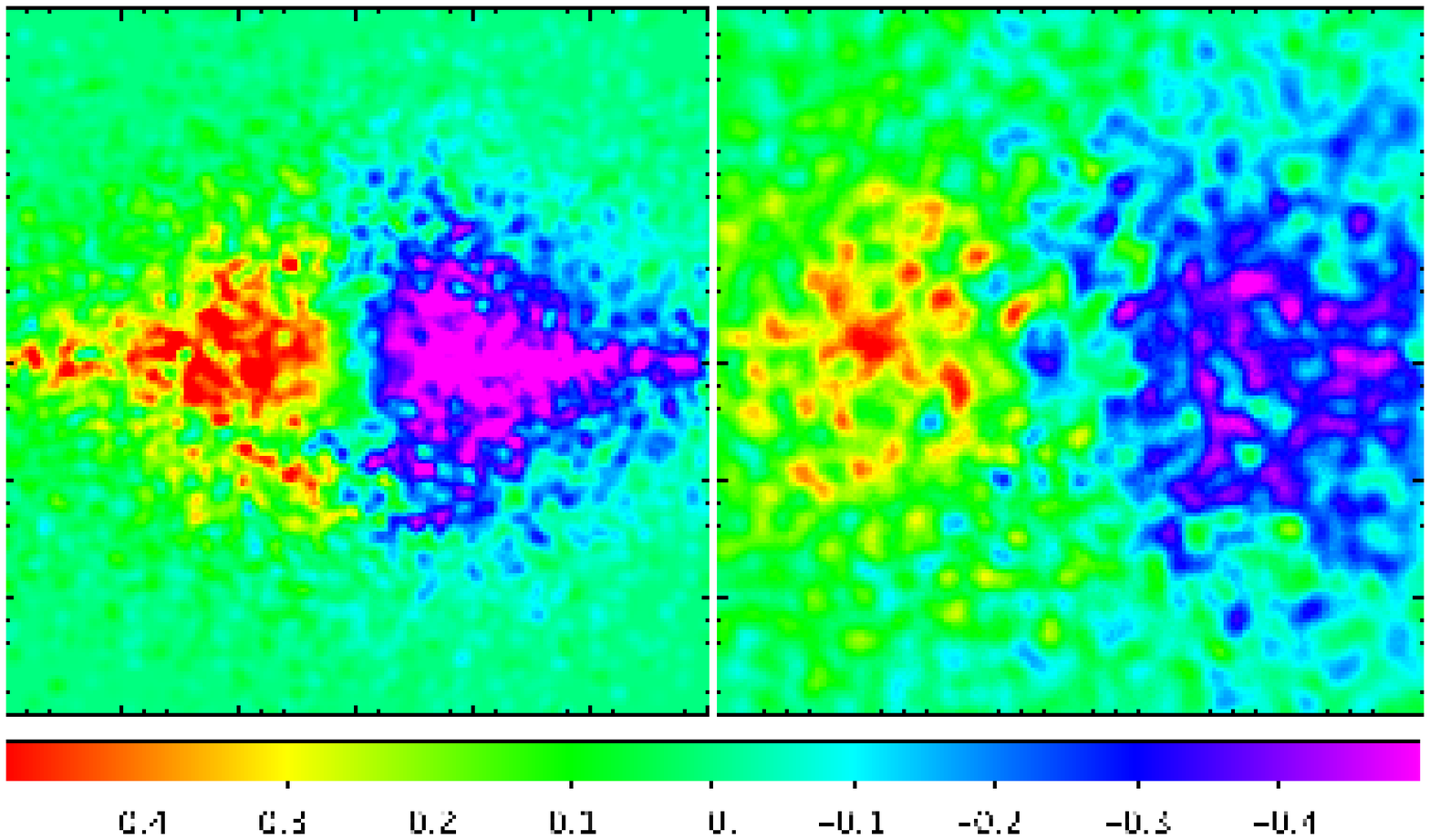}
 \end{minipage}
\begin{minipage}[b]{6cm}
   \centering
  \includegraphics[width=5.8cm,angle=0]{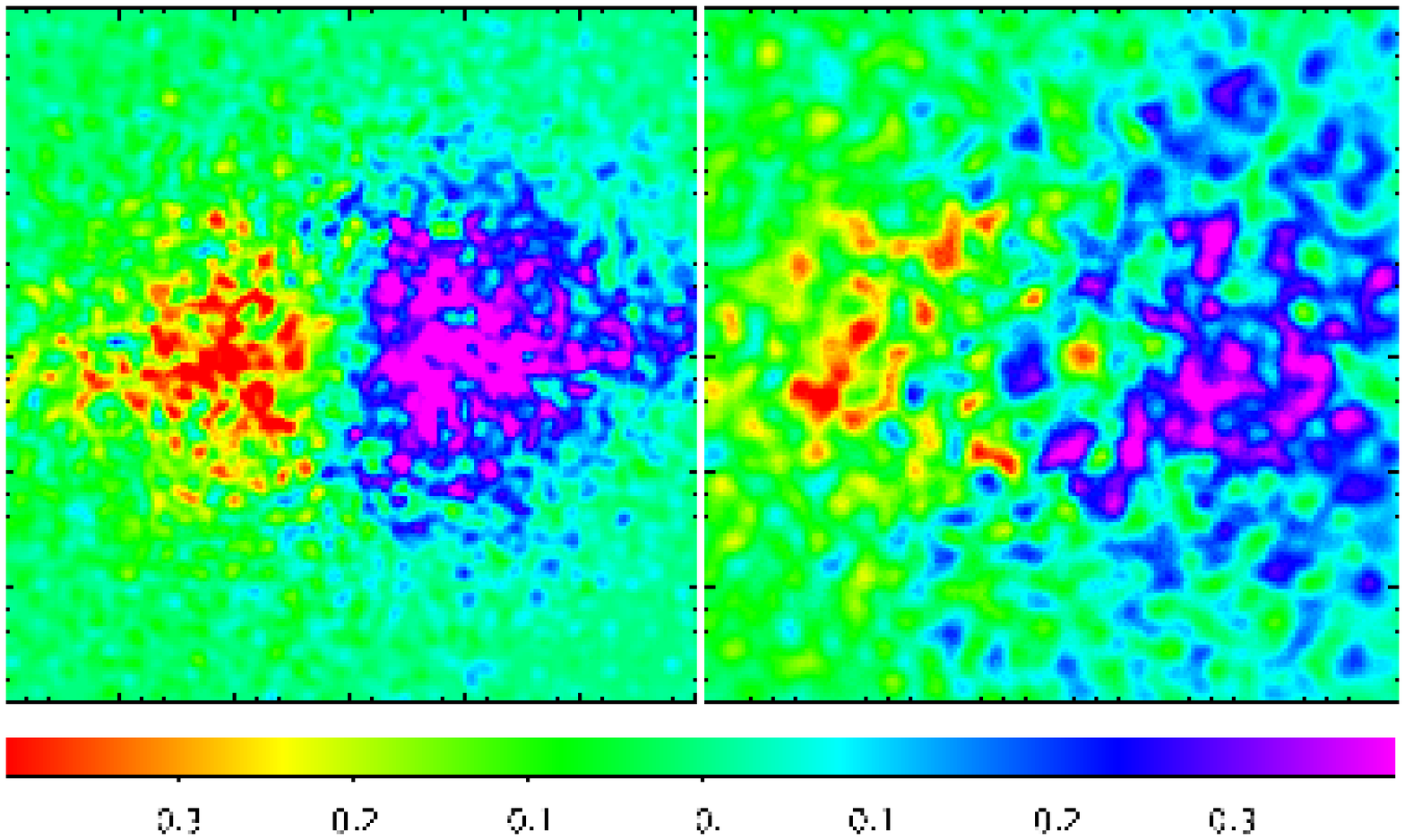}
 \end{minipage}
\caption{2D velocity maps of merger remnants. Each couple of figures refers to a different merger between an elliptical and a spiral galaxy. Two E0-Sa mergers (top panels) and two E0-Sb mergers (bottom panels) are shown, corresponding to encounters with  different initial orbital parameters. For each remnant a large field of view (40 kpc x 40 kpc) and a zoom (10 kpc x 10 kpc) are shown. Velocities are in units of 100 km/s.  \label{maps}}
\end{figure}

\begin{figure}
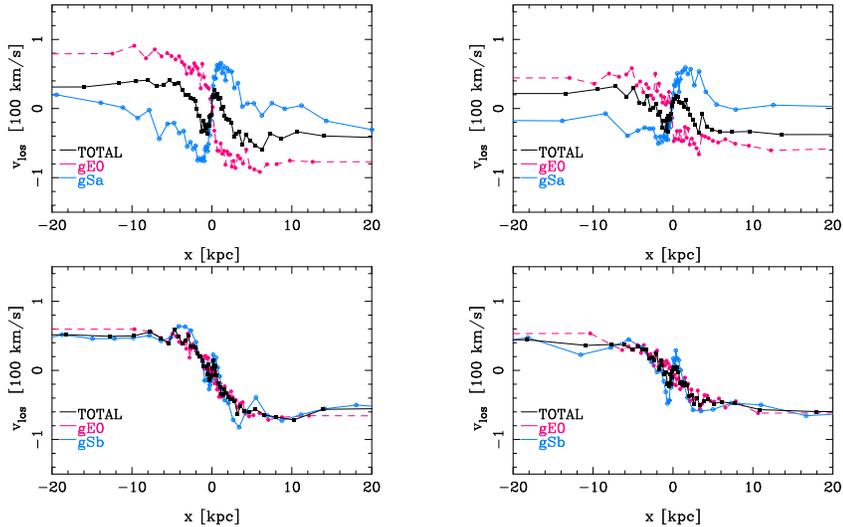

 \begin{minipage}[b]{6.cm}
   \centering
  \includegraphics[width=3.45cm,angle=270]{dimatteo5.ps}
 \end{minipage}
\begin{minipage}[b]{6.cm}
   \centering
  \includegraphics[width=3.45cm,angle=270]{dimatteo6.ps}
 \end{minipage}
\\
\begin{minipage}[b]{6.cm}
   \centering
  \includegraphics[width=3.45cm,angle=270]{dimatteo7.ps}
 \end{minipage}
\begin{minipage}[b]{6.cm}
   \centering
  \includegraphics[width=3.45cm,angle=270]{dimatteo8.ps}
 \end{minipage}
\caption{Rotation curves of remnants of coplanar elliptical-spiral mergers, whose 2D velocity maps are shown in Fig.\ref{maps}. Top panels correspond to E0-Sa mergers, bottom panels to  E0-Sb ones. In each panel three curves are shown: the line-of-sight velocity profile of the old stellar component (black curve), the line-of-sight profile of stars initially belonging to the elliptical (red curve), and that of stars initially belonging to the spiral (blue curve). \label{curves}}
\end{figure}

The 2D velocity maps and the rotation curves of the remnants of retrograde elliptical-spiral mergers are  analyzed at least 400 Myr after the coalescence of the two systems (see Figs.\ref{maps} and \ref{curves}). The rotation curves clearly show that the counter-rotating region is completely associated to stars initially belonging to the spiral galaxy and  that counter-rotating cores are more extended in the case of E0-Sa merger remnants rather than for E0-Sb ones \citep{dimatteo07d}.\\
Initially, the redistribution of the orbital angular momentum affects mostly the outer parts of the galaxies. As the interaction proceeds toward the final merging phase, tidal torques begin to affect inner regions too. If the tidal forces in the inner regions are not strong enough to reverse the initial spin of the spiral galaxy, the spiral continues to rotate with a spin parallel to the initial one, while, in turn, the outer parts can acquire an amount of orbital angular momentum sufficient to reverse their initial rotation. This causes the emergence of a counter-rotating region, among stars initially in the disk galaxy \citep{dimatteo07c}.

\end{document}